# Numerical calculation of the relativistic acceleration of an electron in curved spacetime using the Dirac equation

J.D. Franson

*Physics Dept., University of Maryland at Baltimore County, Baltimore, MD 21250 USA*

The relativistic acceleration of an electron in a uniform gravitational field is calculated numerically using the generalization of the Dirac equation to curved spacetime. Equivalent results are also obtained analytically using an iterative approach that is not based on the WKB approximation, which has been used extensively in the past. The acceleration is found to be proportional to a factor of $(1-3v^2/c^2)$ using Schwarzschild coordinates, which is consistent with the classical geodesic of a particle. These techniques may be useful in resolving the differences between commonly-used approximations.

## I. INTRODUCTION

The Dirac equation for a relativistic electron can be generalized to curved spacetime in a straightforward way [1-3]. Solving the resulting equations is not straightforward, however, and the WKB approximation has been extensively used for that purpose [4-8]. It has previously been suggested that the WKB approximation gives results that are not equivalent to those obtained from the leading term in a power-series expansion of the Foldy-Wouthuysen transformation [9]. Here we calculate the relativistic acceleration of an electron in a uniform gravitational field by numerically solving the general relativistic form of the Dirac equation. This approach does not make any approximations and it is found to be in good agreement with the classical geodesic as well as a new iterative approach for solving the Dirac equation.

This work was motivated in part by the fact that quantum mechanics and general relativity are two of the most fundamental theories in physics, and yet there is no consensus on how these two theories could be unified in a consistent way. As a result, it may be useful to consider the behavior of quantum systems in curved spacetime in the most fundamental way possible and with a minimum of approximations. The results presented here include graphical representations of the motion of the wave packet of an electron in curved spacetime as it approaches relativistic velocities.

An interesting feature of these results is that the acceleration is proportional to a factor of $(1-3v^2/c^2)$ using Schwarzschild coordinates, whereas special relativity would give a factor of $(1-v^2/c^2)^{1/2}$ instead due to the effective mass increase of a particle. This gives rise to an apparent reversal in the sign of the acceleration of gravity for velocities greater than $c/\sqrt{3}$. An unexpected factor of $(1+3\Phi_G)$ also appears in the kinetic energy. These effects are artifacts associated with the arbitrary nature of coordinate systems, as will be discussed in more detail below.

The remainder of this paper begins in Section II with a brief review of the generalization of the Dirac equation to curved spacetime and the form of the resulting equations for the special case of a uniform gravitational field. The acceleration of an electron is calculated analytically in Section III using an iterative approach that is not equivalent to either the WKB approximation [4-8] or the Foldy-Wouthuysen transformation [9-11]. The relativistic wave packet of an electron is then calculated numerically in Section IV and plotted as a function of time. Both the numerical and analytic calculations are consistent with the classical geodesic of a particle. A summary and conclusions are presented in Section V, while several appendices describe the covariance of the Dirac equation, the connection with the Foldy-Wouthuysen transformation, and additional details of the numerical and analytic calculations.

## II. DIRAC EQUATION IN CURVED SPACETIME

In its simplest form, the weak equivalence principle states that all objects must fall at the same rate in a uniform gravitational field regardless of their composition [12,13]. The formulation of physical theories in curved spacetime also requires the strong equivalence principle, which states that the laws of physics must reduce to those of flat spacetime (special relativity) in a local freely-falling reference frame [12].

The Dirac equation in the flat (Minkowski) spacetime of special relativity can be written as

$$i\gamma_{(0)}{}^\mu \partial_\mu \psi(x) = m\psi(x), \qquad (1)$$

where $m$ is the mass of an electron [14]. A system of units with $c = \hbar = 1$ will be used for simplicity, where $c$ is the speed of light and $\hbar$ is Planck's constant divided by $2\pi$. (Those constants will occasionally be reinserted for clarity.) The index $\mu$ takes on integer values from 0 to 3, which correspond to the four dimensions of spacetime. A sum over $\mu$ is performed in accordance with Einstein's summation convention for repeated indices. The operator $\partial_\mu = \partial / \partial x^\mu$ while $\psi$ is a four-component spinor (a column vector of complex numbers). The $\gamma_{(0)}{}^\mu$ are the set of $4 \times 4$ Dirac matrices [14], where the subscript (0) indicates that these are the constant matrices used in the original Dirac equation in flat spacetime.

In the flat spacetime of special relativity, the $\gamma_{(0)}{}^\mu$ are required to satisfy the condition [14]

$$\gamma_{(0)}{}^\mu \gamma_{(0)}{}^\nu + \gamma_{(0)}{}^\nu \gamma_{(0)}{}^\mu = 2\eta^{\mu\nu}. \tag{2}$$

Here $\eta^{\mu\nu}$ is the metric of flat spacetime, which is a diagonal matrix with the elements $\{1, -1, -1, -1\}$. Eq. (2) is sufficient to ensure that the Dirac equation is covariant under a Lorentz transformation.

The Dirac equation can be generalized to a curved spacetime using a series of freely-falling inertial reference frames as described in more detail in Appendix A. The result [1-3] is that Eq. (1) must be replaced with

$$i\gamma^\mu(x) \mathcal{D}_\mu \psi(x) = m\psi(x), \tag{3}$$

where the new set of matrices $\gamma^\mu(x)$ are now dependent on the coordinates $x^\mu$. Eq. (2) is also replaced with

$$\gamma^\mu(x)\gamma^\nu(x) + \gamma^\nu(x)\gamma^\mu(x) = 2g^{\mu\nu}(x), \tag{4}$$

where $g^{\mu\nu}(x)$ is the metric of curved spacetime. The covariant derivative $\mathcal{D}_\mu$ is defined [12] by

$$\mathcal{D}_\mu \equiv \partial_\mu + \Gamma_\mu. \tag{5}$$

The additional spin connection term $\Gamma_\mu$ in Eq. (5) is chosen to ensure that $\mathcal{D}_\mu$ transforms as a vector under Lorentz transformations of the freely-falling reference frames as described in Appendix A.

We will consider a uniform gravitational field in which the force $\mathbf{f}_G$ of gravity on a test mass $m$ is independent of position. This can be approximated to arbitrary accuracy by taking the system of interest to be located a large distance $R$ from a mass M as illustrated in Fig. 1. In the limit of large $R$, $\mathbf{f}_G$ is given by

$$\mathbf{f}_G = -G\frac{Mm}{R_0^2}\hat{e}_z \tag{6}$$

where $G$ is the gravitational constant, $R_0$ is the distance to the center of the coordinate system defined in Fig. 1, and $\hat{e}_z$ is a unit vector along the z axis. Any departures from Eq. (6) are of the order of $1/R_0^3$ and negligible in this limiting case.

Under these conditions, the Schwarzschild metric reduces [15] to

$$g^{\mu\nu} = \begin{bmatrix} (1-2\Phi_G) & 0 & 0 & 0 \\ 0 & -(1+2\Phi_G) & & \\ & & -(1+2\Phi_G) & \\ & & & -(1+2\Phi_G) \end{bmatrix} \tag{7}$$

where $\Phi_G(z) = -GM/(R_0 + z)$. Here we have assumed that $|\Phi_G(z)|/c^2 \ll 1$ and Eq. (7) only includes terms that are first order in $\Phi_G(z)$ as will be done throughout the rest of the paper. The spatial dependence of $\Phi_G(z)$ has not been shown in Eq. (7) in order to simplify the notation.

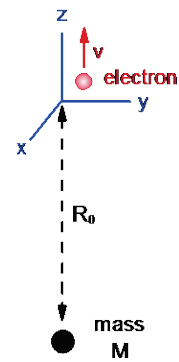

FIG. 1. A uniform gravitational field created by a mass $M$ located a large distance $R_0$ from a coordinate system labelled x, y, z. An electron moving with velocity $v$ in the z direction is located near the center of the coordinate system.

For the diagonal metric of Eq. (7), the $\gamma^\mu(x)$ matrices that satisfy Eq. (4) are given by

$$\begin{aligned}\gamma^0(z) &= [1 - \Phi_G(z)]\gamma_{(0)}{}^0 \\ \gamma^i(z) &= [1 + \Phi_G(z)]\gamma_{(0)}{}^i \quad (i=1,3)\end{aligned} \tag{8}$$

to first order in $\Phi_G$. The origin of the factors of $[1 \pm \Phi_G(z)]$ can be understood from a more fundamental point of view using the vierbein formalism.

The spin connection $\Gamma_\mu$ does not produce any acceleration in a uniform gravitational field as shown in Appendix D. The $\Gamma_\mu$ term in the Hamiltonian is a factor of $(v/c)(\lambda_C/R_0)$ smaller than the usual $m\Phi_G(z)$ term, where $\lambda_C$ is the Compton wavelength of the electron. The $\Gamma_\mu$ term in Eq. (5) will be neglected in the analytic calculations for both of these reasons, although it will be included in the numerical calculations for comparison purposes. The $\Gamma_\mu$ term would produce a small spin-dependent acceleration [4-9] in a spatially-varying gravitational field that is analogous to the acceleration of a rotating classical object.

As usual [14,16], it will be convenient to introduce another set of matrices $\boldsymbol{\alpha}_{(0)}$ and $\beta_{(0)}$ that satisfy the condition

$$\gamma_{(0)}{}^0 = \beta_{(0)} \qquad (9)$$
$$\gamma_{(0)}{}^i = \beta_{(0)} \alpha_{(0)}{}^i.$$

Combining Eqs. (3), (5), (8), and (9) allows the Dirac equation in curved spacetime to be rewritten under these conditions as

$$i\frac{\partial \psi}{\partial t} = (1+2\Phi_G)\boldsymbol{\alpha}_{(0)} \cdot \left(\frac{1}{i}\nabla\right)\psi + (1+\Phi_G)\beta_{(0)}m\psi. \quad (10)$$

Eq. (10) is the Dirac equation in curved spacetime to first order in $\Phi_G$ for a uniform gravitational field, neglecting the spin-dependent $\Gamma_\mu$ term. This equation is the same as the Dirac equation of special relativity aside from the $\Phi_G$ terms. Since $\beta_{(0)}$ is a diagonal matrix with elements of $\pm 1$, the last term in Eq. (10) corresponds to a gravitational potential of $m\Phi_G$ for an electron, which is analogous to the gravitational potential in the nonrelativistic Schrodinger equation.

It can be seen that the effects of gravity enter into the Dirac equation through the metric rather than by including them directly in the Hamiltonian or Lagrangian. One consequence of this is the factor of $(1+2\Phi_G)$ that multiplies the kinetic energy term in Eq. (10). The effects of this term will be investigated in the remainder of the paper.

## III. ITERATIVE CALCULATION OF THE ACCELERATION

The acceleration of an electron in a uniform gravitational field will now be calculated for the case in which its velocity $v$ along the z axis is small compared to the speed of light. The result will be derived analytically to order $v^2/c^2$ while the corresponding calculations will be performed numerically in the following section with no assumptions other than $\Phi_G/c^2 \ll 1$.

As is often the case [16], it will be convenient to write the four-component spinor $\psi(x^\mu)$ as a combination of two two-component spinors $\phi(x^\mu)$ and $\chi(x^\mu)$:

$$\psi(x^\mu) = \begin{bmatrix} \phi_1(x^\mu) \\ \phi_2(x^\mu) \\ \chi_1(x^\mu) \\ \chi_2(x^\mu) \end{bmatrix}.$$

(11)

Inserting Eq. (11) into Eq. (10) allows the Dirac equation to be written in the equivalent form

$$i\frac{\partial}{\partial t}\phi = (1+2\Phi_G)\boldsymbol{\tau}\cdot\left(\frac{1}{i}\nabla\right)\chi + (1+\Phi_G)m\phi$$
$$i\frac{\partial}{\partial t}\chi = (1+2\Phi_G)\boldsymbol{\tau}\cdot\left(\frac{1}{i}\nabla\right)\phi - (1+\Phi_G)m\chi. \quad (12)$$

Here $\boldsymbol{\tau}$ is the set of $2\times 2$ Pauli spin matrices. Eq. (12) is often used in special relativity as well, the only difference being that $\Phi_G = 0$ in that case.

Eq. (12) can be put into a form that is similar to the non-relativistic Schrodinger equation using an iterative approach [16] that makes use of the fact that $\chi \sim (v/c)\phi$. This allows $\chi$ to be written in terms of $\phi$ to any desired order of $v/c$ and eliminated from the equations. To fourth-order in $v/c$ this gives

$$i\frac{\partial \phi}{\partial t} = \frac{1}{2m}\hat{\mathbf{p}}\cdot\left[(1+3\Phi_G)\hat{\mathbf{p}}\right]\phi + (1+\Phi_G)m\phi - \frac{1}{8m^3}\hat{\mathbf{p}}^4\phi. \quad (13)$$

as is shown in more detail in Appendix B. Here the symbol $\hat{\mathbf{p}}$ is defined as $\hat{\mathbf{p}} \equiv \nabla/i$, although it may not correspond exactly to the momentum. A spin-dependent term proportional to $\boldsymbol{\tau}\cdot(\nabla\Phi_G \times \hat{\mathbf{p}})\phi$ has been omitted from Eq. (13) since it is zero for a velocity along the z axis as is assumed in Fig. 1. Similar results can also be obtained using a Foldy-Wouthuysen transformation [9-11] as described in Appendix E.

In the Heisenberg picture [16] the rate of change of the position is given by

$$\frac{d\hat{x}}{dt} \equiv \hat{v} = \frac{1}{i}[\hat{x}, \hat{H}]. \quad (14)$$

Here the Hamiltonian $\hat{H}$ corresponds to the operators on the right-hand side of Eq. (13) while $\hat{x}$ is the position operator.

The acceleration $\hat{a}$ of the particle is similarly given by

$$\hat{a} = \frac{d\hat{v}}{dt} = \frac{1}{i}[\hat{v}, \hat{H}]. \tag{15}$$

As shown in Appendix B, these commutators can be evaluated to give

$$\hat{a} = \frac{d^2\hat{z}}{dt^2} = -\nabla\Phi_G\left(1 - 3\frac{\hat{p}_z^2}{m^2}\right). \tag{16}$$

If we replace the operators with their expectation values, then to first order in $\nabla\Phi_G$

$$\frac{d^2z}{dt^2} = -\nabla\Phi_G\left(1 - 3\frac{v_z^2}{c^2}\right). \tag{17}$$

For simplicity, expectation values such as $\langle\hat{v}^2\rangle$ have been written as $v^2$. The transition from Eq. (16) to Eq. (17) makes use of the fact that any dependence of $\hat{v}$ on $\Phi_G$ would give effects that are second-order in $\Phi_G$. Eq. (17) is one of the main results of this paper.

The acceleration in eq. (17) has the interesting feature that it is dependent on a factor of $(1 - 3v^2/c^2)$, whereas a dependence on $(1 - v^2/c^2)^{1/2}$ would be expected in special relativity. This is closely related to the factor of $(1 + 3\Phi_G)$ that appears in the kinetic energy term in Eq. (13).

For comparison, a classical particle undergoing free fall in a uniform gravitational field will follow a geodesic curve. In Schwarzschild coordinates, that corresponds to an acceleration $a_g$ given to first order in $\Phi_G$ by

$$a_g = -\nabla\Phi_G\left(1 - \frac{3v_z^2}{c^2}\right), \tag{18}$$

as is derived in Appendix F. It can be seen from Eqs. (17) and (18) that the velocity-dependent term in the acceleration of an electron as given by the Dirac equation is the same as that of a classical particle following a geodesic, at least to order $v^2/c^2$. This dependence of the acceleration on a factor of $(1 - 3v^2/c^2)$ is due to the arbitrary nature of the choice of coordinate system, as will be discussed later in the paper.

This iterative approach can be repeated to obtain higher-order terms in the correction to the acceleration, although the process becomes increasingly tedious.

## IV. NUMERICAL CALCULATIONS

The WKB approximation, power-series expansions using the Foldy-Wouthuysen transformation, and the iterative approach described above are all approximate in nature. In this section, we describe the results of calculations in which the Dirac equation was integrated numerically for a range of initial velocities using the full Dirac equation (12) with the $\Gamma_\mu$ term included. As a result, the numerical calculations make no assumptions other than $|\Phi_G|/c^2 \ll 1$. Aside from verifying the results of the analytic calculations, the numerical calculations provide graphical representations of the wave function of an electron moving at relativistic velocities in curved spacetime.

As illustrated in Fig. 1, the gravitational field was chosen to lie along the negative $z$ axis. For a uniform gravitational field, the gradient $\nabla\Phi_G$ is a constant while $\Phi_G$ is given by

$$\Phi_G(z) = \Phi_{G0} + z\nabla\Phi_G. \tag{19}$$

Here $\Phi_{G0}$ is the gravitational potential at the center of the coordinate system. The initial wave function was chosen to be independent of the coordinates $x$ and $y$ in order to reduce the computational burden. Eq. (19) was then inserted into Eq. (12) and $\psi(z,t)$ was calculated numerically using Mathematica as described in more detail in Appendix C.

The acceleration in a uniform gravitational field is independent of the spin and we considered the case in which the spin was oriented along the positive z axis. In that case $\phi_2 = \chi_2 = 0$ and it was only necessary to calculate the two nonzero components $\phi_1(z,t)$ and $\chi_1(z,t)$.

Fig. 2 shows the real parts of $\phi_1(z,t)$ and $\chi_1(z,t)$ along with the density $\rho(z,t) \equiv \psi^\dagger(z,t)\psi(z,t)$ for the case in which the gravitational field is zero. The data were plotted over a relatively short time interval here in order to show the oscillations in $\phi_1(z,t)$ and $\chi_1(z,t)$. Despite those oscillations, it can be seen that the density $\rho(z,t)$ corresponds to a smooth Gaussian wave packet that moves at a constant velocity.

Fig. 3 shows the density $\rho(z,t)$ over a much longer time interval where the gravitational field can produce a significant change in the velocity of the wave packet. Fig. 3a corresponds to zero gravitational field as before while Figs. 3b and 3c correspond to $\nabla\Phi_G = \pm 0.05$ in units where $c = \hbar = 1$. The acceleration is independent of the mass of the electron, which was arbitrarily taken to be $m = 100$ in these units in order to reduce the computational burden associated with the rapid oscillation of the phase. It can be seen that the wave packet as a whole is accelerated much as it would be classically.

The displacement of the final position of the wave packet was used to obtain a numerical estimate of the acceleration of an electron as a function of its velocity. The results of the numerical calculations are compared with the analytic calculations in Fig. 4. The acceleration of gravity is independent of a particle's velocity in Newtonian physics, as illustrated by the dashed black line. The results obtained analytically from the generalization of the Dirac equation to curved spacetime [1-3] are shown by the solid red line, as given by Eq. (17). Finally, the results of the numerical calculations are indicated by the blue plus signs. The numerical estimates are correct to approximately four significant digits and any error bars would be too small to include in the figure.

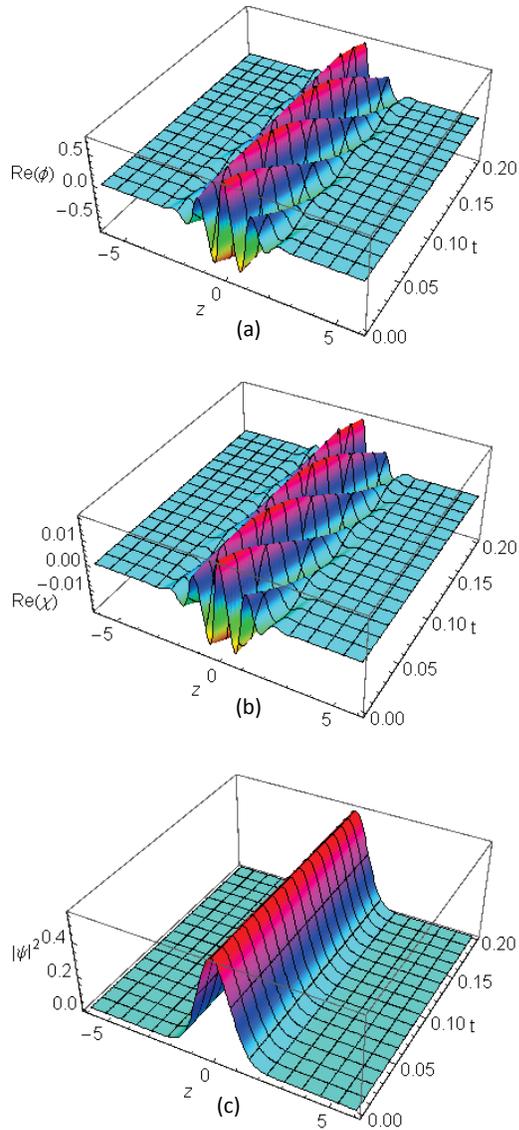

FIG. 2. Numerical solution of the Dirac equation over a relatively short period of time with zero gravitational field. (a) Real part of the $\phi_1(z,t)$ component as a function of position $z$ and time $t$. (b) Real part of the $\chi_1(z,t)$ component. (c) The density $\psi^\dagger\psi$. These plots correspond to a velocity of $v/c = 0.05$. (Arbitrary units with $c = \hbar = 1$.)

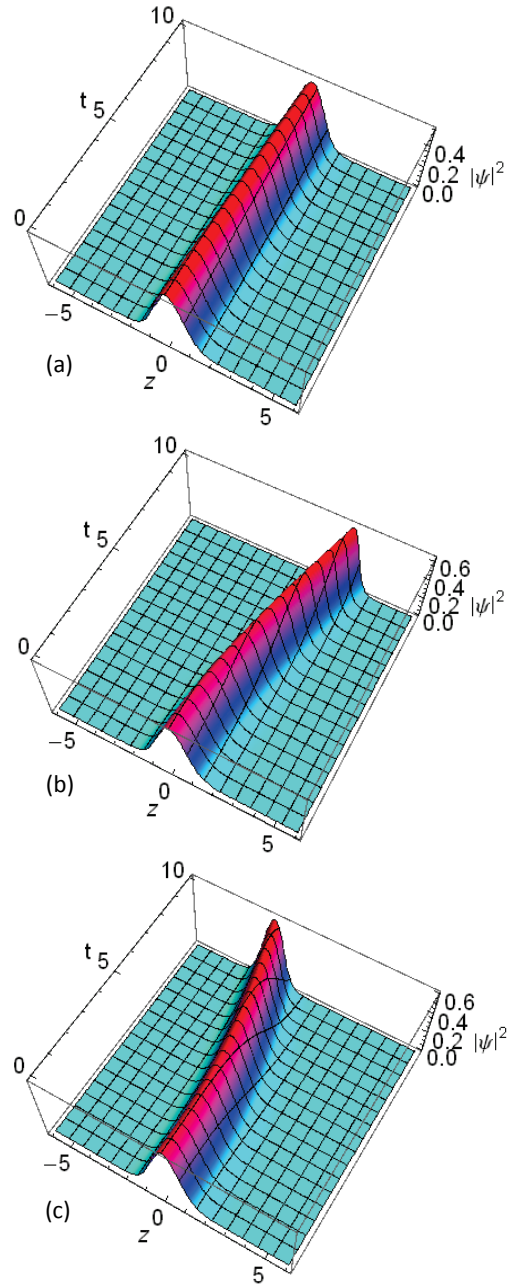

FIG. 3. Numerical solution of the Dirac equation showing the motion of a wave packet accelerated by a gravitational field. The density $\rho(z,t) \equiv |\psi(z,t)|^2$ is plotted as a function of position $z$ and time $t$. (a) $\nabla\Phi_G = 0$, no gravitational field. (b) $\nabla\Phi_G = -0.05$. (c) $\nabla\Phi_G = 0.05$. (Arbitrary units with $c = \hbar = 1$.)

It can be seen that the analytic and numerical calculations agree to better than 1%. The small difference between the two is due to the fact that the analytic calculations are only correct to order $v^2/c^2$. The analytic calculations for a uniform gravitational field are equivalent to the geodesic trajectory of a classical particle as can be seen from Eqs. (17) and (18).

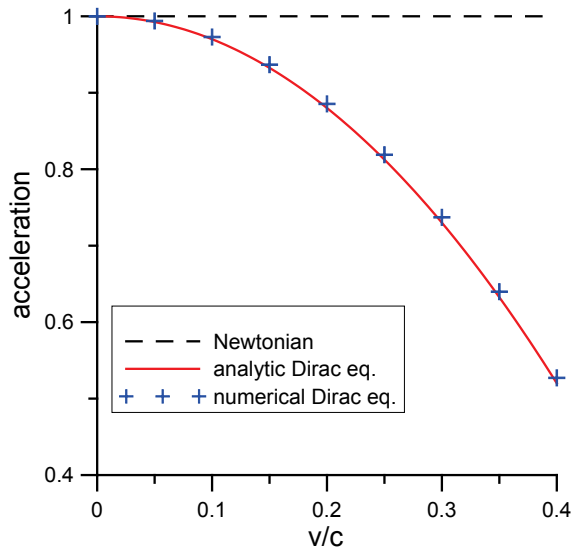

FIG. 4. Comparison of the acceleration of an electron in a uniform gravitational field as a function of its velocity. It can be seen that the analytic and numerical calculations based on the Dirac equation are in good agreement.

## V. SUMMARY AND CONCLUSIONS

The quantum-mechanical motion of a relativistic electron in curved spacetime is usually analyzed using the WKB approximation [4-8]. It has been suggested that a power-series expansion based on the Foldy-Wouthuysen transformation may give different results than the WKB approximation for the spin dependence of the acceleration [9]. Thus there is a need to be able to resolve these differences in the results from various approximation techniques.

Here we have described a numerical technique that can be used to calculate the motion of the wave packet of an electron directly from the form of the Dirac equation generalized to curved spacetime. For simplicity, we have considered the case of a uniform gravitational field to first order in $\Phi_G/c^2$, but the same technique can also be used for the more general case. It was found that the relativistic acceleration of an electron in a uniform gravitational field corresponds to that of a classical geodesic as would be expected.

An iterative approach for calculating the acceleration of an electron in curved spacetime was also described. This approach is similar to techniques commonly used in special relativity [16] and it has the advantage of being relatively straightforward. The acceleration calculated using this iterative approach was in good agreement with the numerical calculations.

The acceleration was found to be proportional to a factor of $(1-3v^2/c^2)$ in Schwarzschild coordinates, whereas special relativity would give a factor of $(1-v^2/c^2)^{1/2}$ instead for the relativistic acceleration of a particle. This result is closely related to the factor of $(1+3\Phi_G)$ that appears in the kinetic energy term of Eq. (13), whereas one might have intuitively expected a factor of $(1+\Phi_G)$ as a correction to the energy/mass of the system instead. Both of these effects are artifacts related to the arbitrary nature of the choice of a coordinate system, as can be seen from the fact that the acceleration reverses sign for velocities greater than $c/\sqrt{3}$. This does not mean that the force of gravity will be reversed at high velocities. Instead, it reflects the fact that the coordinate $t$ is arbitrary to some degree and is not equivalent to the proper time $\tau$. This leads to an acceleration that is correct when measured using Schwarzschild coordinates but with the potential to be misleading.

In addition to providing a way to check the accuracy of various approximation techniques, the numerical results presented here provide additional physical insight into the evolution of quantum systems in curved spacetime.

## ACKNOWLEDGEMENTS

The author thanks R.Y. Chiao, M.S. Goodman, and T.B. Pittman for their comments on the manuscript.

## APPENDIX A: COVARIANCE AND THE DIRAC EQUATION

The Dirac equation can be generalized to curved spacetime based on the requirement that it be covariant. The approach will closely follow the procedure given in Weinberg's text [12] for deriving the generally-covariant form of an arbitrary theory. The Dirac equation can be written in several different forms [1], all of which are equivalent to the description used here.

We will first consider the situation in special relativity, where a homogeneous Lorentz transformation from coordinates $x$ to a new set of coordinates $x'$ is given by

$$x'^\mu = \Lambda^\mu_{\ \nu} x^\nu, \tag{A1}$$

where the coefficients $\Lambda^\mu{}_\nu$ reflect any rotations and boosts. The Lorentz transformations form a group in the sense that the product of any two Lorentz transformations $\Lambda_2$ and $\Lambda_1$ is another Lorentz transformation $\Lambda_3$:

$$\Lambda_3 = \Lambda_2 \Lambda_1. \quad (A2)$$

If we consider a quantum system described by a wave function (or field) $\psi_m$ with $M$ components, then under a Lorentz transformation the new wave function is assumed to be transformed by a $M \times M$ matrix $S(\Lambda)$:

$$\psi'_m = \sum_{n=1}^{M} [S(\Lambda)]_{mn} \psi_n. \quad (A3)$$

Here $S(\Lambda)$ is an irreducible matrix representation of the Lorentz group, which is to say that the matrices $S(\Lambda)$ satisfy the same group multiplication law as $\Lambda$ itself:

$$S(\Lambda_3) = S(\Lambda_2) S(\Lambda_1). \quad (A4)$$

The required properties of the $S(\Lambda)$ can be found by considering an infinitesimal Lorentz transformation given by

$$\Lambda^\mu{}_\nu = \delta^\mu{}_\nu + \omega^\mu{}_\nu. \quad (A5)$$

Here $\delta^\mu{}_\nu$ is the Kronecker delta function and $\omega^\mu{}_\nu$ is a set of infinitesimal parameters that once again describe any rotations or boosts in the transformation. For an infinitesimal Lorentz transformation of this kind, the most general form of $S(\Lambda)$ is given by

$$S(1+\omega) = 1 + \frac{1}{2} \omega^{\alpha\beta} \sigma_{\alpha\beta}, \quad (A6)$$

where each of the $\sigma_{\alpha\beta}$ is a $M \times M$ matrix. Eq. (A6) can be thought of as the first term in a Taylor series expansion.

By inserting Eq. (A6) into Eq. (A4), it can be shown [12] that the $\sigma_{\alpha\beta}$ must satisfy a set of commutation relations given by

$$[\sigma_{\alpha\beta}, \sigma_{\gamma\delta}] = \eta_{\gamma\beta} \sigma_{\alpha\delta} - \eta_{\gamma\alpha} \sigma_{\beta\delta} + \eta_{\delta\beta} \sigma_{\gamma\alpha} - \eta_{\delta\alpha} \sigma_{\gamma\beta}. \quad (A7)$$

Group theory techniques can be used to find explicit representations of the $\sigma_{\alpha\beta}$ that satisfy Eq. (A7). For the Dirac theory, $S(\Lambda)$ and the $\sigma_{\alpha\beta}$ are $4 \times 4$ matrices where the $\sigma_{\alpha\beta}$ are given [14] by

$$\sigma^{\mu\nu} = \frac{1}{4}[\gamma_{(0)}{}^\mu, \gamma_{(0)}{}^\nu] = \frac{1}{4}\left(\gamma_{(0)}{}^\mu \gamma_{(0)}{}^\nu - \gamma_{(0)}{}^\nu \gamma_{(0)}{}^\mu\right). \quad (A8)$$

The covariance of the Dirac equation under a Lorentz transformation in special relativity can now be demonstrated using the fact that the $\gamma_{(0)}{}^\mu$ satisfy the condition [14]

$$S(\Lambda)^{-1} \gamma_{(0)}{}^\mu S(\Lambda) = \Lambda^\mu{}_\nu \gamma_{(0)}{}^\nu. \quad (A9)$$

If the Dirac equation (1) is satisfied in the $x$ coordinate system, then we can transform it into the $x'$ coordinate system using

$$\begin{aligned} \psi &= D^{-1} \psi' \\ \partial_\mu &= \left(\Lambda^{-1}\right)^\nu{}_\mu \partial'_\nu. \end{aligned} \quad (A10)$$

Inserting Eqs. (A9) and (A10) into the Dirac Eq. (1) and using the fact that

$$\left(\Lambda^{-1}\right)^\nu{}_\mu = \Lambda_\mu{}^\nu \quad (A11)$$

gives

$$i\gamma_{(0)}{}^\mu \partial'_\mu \psi' = m\psi'. \quad (A12)$$

Here both sides of the equation were multiplied by $S(\Lambda)$ in which case all the factors of $S(\Lambda)$ and $S^{-1}(\Lambda)$ cancel out. Eq. (A12) corresponds to the Dirac equation in the new coordinate system as desired.

This approach can be generalized to the curved spacetime of general relativity using the vierbein formalism [12]. At any point $X$ in a global coordinate system $x^\mu$, it is possible to construct a local reference frame $\xi_X{}^\alpha$ where the metric $g_{\mu\nu} = \eta_{\mu\nu}$ in the vicinity of $X$. The vierbein $V^\alpha{}_\beta$ are defined by

$$V^\alpha{}_\mu(X) = \left(\frac{\partial \xi_X{}^\alpha(x)}{\partial x^\mu}\right)_{x=X}. \quad (A13)$$

The Dirac equation (and other physical theories) can then be put into a covariant form by contracting all vectors and tensors with $V^\alpha{}_\beta$ to produce a scalar under transformations of the global coordinates $x^\mu$. For example, given a vector $A^\mu$ in the global coordinate system, we can define a set of four scalars $*A^\alpha$ given by

$$*A^\alpha = V^\alpha{}_\mu A^\mu. \tag{A14}$$

The covariant derivative $\mathcal{D}_\alpha$ is then chosen in such a way that, under a Lorentz transformation of the local inertial frame $\xi_X{}^\alpha$,

$$\mathcal{D}_\alpha *\psi(x) \to \Lambda_\alpha{}^\beta(x) S(\Lambda(x)) \mathcal{D}_\beta *\psi(x). \tag{A15}$$

This requires that

$$\mathcal{D}_\alpha = V_\alpha{}^\mu \left[ \frac{\partial}{\partial x^\mu} + \Gamma_\mu \right], \tag{A16}$$

where the spin connection $\Gamma_\mu$ can be shown [12] to be given by

$$\Gamma_\mu = \frac{1}{2} \sigma^{\alpha\beta} V_\alpha{}^\nu V_{\beta\nu;\mu}. \tag{A17}$$

The semicolon in Eq. (A17) denotes the usual covariant derivative of a vector using the Christoffel symbol, which is necessary to ensure that the resulting theory is covariant under an arbitrary change of the global coordinates $x^\mu$.

The covariant form of the Dirac equation can now be written as

$$i\gamma_{(0)}{}^\alpha \mathcal{D}_\alpha *\psi = m*\psi. \tag{A18}$$

The covariance of the theory under Lorentz transformations of the local inertial reference frames can be demonstrated using Eq. (A15) in the same way as was done for flat spacetime. Eq. (A18) can be put into the same form as Eq. (3) by defining [2]

$$\gamma^\alpha(x) = \gamma_{(0)}{}^\beta V_\beta{}^\alpha, \tag{A19}$$

which can be shown to be consistent with Eq. (4). The wave function is the same in all global coordinate systems and $*\psi = \psi$.

The approach outlined above maintains the covariance of the Dirac theory under Lorentz transformations of the local inertial reference frames as well as arbitrary transformations of the global coordinate system. It also ensures that the theory will reduce to the Dirac equation of special relativity as $g^{\mu\nu} \to \eta^{\mu\nu}$.

**APPENDIX B: DETAILS OF THE ANALYTIC CALCULATIONS**

Equation (12) is valid for the Schwarzschild metric of Eq. (7) to first order in $\Phi_G$ and neglecting the spin-dependent $\Gamma_\mu$ term. The second line of Eq. (12) can be rewritten in the form

$$\chi = \frac{(1+2\Phi_G)}{2m}(\boldsymbol{\tau}\cdot\hat{\mathbf{p}})\phi - \frac{1}{2m}\left(i\frac{\partial}{\partial t} - m + m\Phi_G\right)\chi, \tag{B1}$$

where $\hat{\mathbf{p}} \equiv \frac{1}{i}\nabla$ as before. Using the fact that $|\chi| << |\phi|$ for $v/c << 1$ allows Eq. (B1) to be rewritten as

$$\chi = \frac{(1+2\Phi_G)}{2m}(\boldsymbol{\tau}\cdot\hat{\mathbf{p}})\phi \tag{B2}$$

to first order in $v/c$.

We can obtain an expression for $\chi$ that is accurate to second order in $v/c$ by inserting Eq. (B2) back into Eq. (B1), which gives

$$\chi = \frac{(1+2\Phi_G)}{2m}(\boldsymbol{\tau}\cdot\hat{\mathbf{p}})\phi - \frac{1}{2m}\left(i\frac{\partial}{\partial t} - m + m\Phi_G\right) \\ \times \frac{(1+2\Phi_G)}{2m}(\boldsymbol{\tau}\cdot\hat{\mathbf{p}})\phi. \tag{B3}$$

A similar iterative procedure is commonly used for the Dirac equation in special relativity [16].

Inserting Eq. (B3) into the first line of Eq. (12) gives

$$i\frac{\partial}{\partial t}\phi = (1+2\Phi_G)(\boldsymbol{\tau}\cdot\hat{\mathbf{p}})\frac{(1+2\Phi_G)}{2m}(\boldsymbol{\tau}\cdot\hat{\mathbf{p}})\phi + m\phi + m\Phi_G\phi \\ -(1+2\Phi_G)(\boldsymbol{\tau}\cdot\hat{\mathbf{p}})\frac{1}{2m}\left(i\frac{\partial}{\partial t} - m + m\Phi_G\right)\frac{(1+2\Phi_G)}{2m}(\boldsymbol{\tau}\cdot\hat{\mathbf{p}})\phi \tag{B4}$$

Eq. (B4) can be simplified by moving $\hat{\mathbf{p}}$ to the left using $\Phi_G(z)\hat{\mathbf{p}} = \hat{\mathbf{p}}\Phi_G(z) + [\Phi_G(z),\hat{\mathbf{p}}]$ until the term with the time derivative on the right-hand side of Eq. (B4) acts directly on $\phi$. We can then make the substitution [16]

$$\left(i\frac{\partial}{\partial t} - m + m\Phi_G\right)\phi = \left(\frac{\hat{p}^2}{2m} + 2m\Phi_G\right)\phi, \tag{B5}$$

which is correct to second-order in $v/c$. Eq. (B4) can be further simplified using the identity

$$(\boldsymbol{\tau}\cdot\mathbf{A})(\boldsymbol{\tau}\cdot\mathbf{B}) = \mathbf{A}\cdot\mathbf{B} + i\boldsymbol{\tau}\cdot(\mathbf{A}\times\mathbf{B}), \tag{B6}$$

where $\mathbf{A}$ and $\mathbf{B}$ are any two vectors.

The result is the Schrodinger-like Eq. (13) where higher-order terms such as $\hat{p}^4\Phi_G\phi$ have been dropped along with $\boldsymbol{\tau}\cdot(\nabla\Phi_G\times\mathbf{p})\phi=0$. Non-Hermitian terms proportional to $i\nabla\Phi_G\cdot\hat{\mathbf{p}}$ also appear in the derivation; these are analogous to the terms involving $i\mathbf{E}\cdot\mathbf{p}$ that appear in special relativity [16] for an electron in an electric field $\mathbf{E}$. In both cases, these non-Hermitian terms are due to the fact that the density associated with $\chi$ changes as the particle accelerates, which it is not reflected in $\phi^\dagger\phi$. The non-Hermitian terms can be removed by renormalizing the wave function with the transformation

$$\phi' = \left(1 + \frac{5\hat{\mathbf{p}}^2}{8m^2}\right)\phi. \qquad (B7)$$

Dropping the primes gives Eq. (13). Similar results can be also obtained using the Foldy-Wouthuysen transformation [9] as is described in a subsequent section.

Commutator techniques can now be used to calculate the acceleration of an electron in a uniform gravitational field. Here we are only interested in the acceleration in the z direction, which is given in the Heisenberg picture by

$$\dot{\hat{z}} = \frac{1}{i}[\hat{z}, \hat{H}], \qquad (B8)$$

where a dot above a symbol indicates the time derivative. The only nonzero commutators in the Hamiltonian of Eq. (13) give

$$\dot{\hat{z}} = \frac{1}{i}[\hat{z}, \frac{1}{2m}\hat{\mathbf{p}}\cdot(1+3\Phi_G)\hat{\mathbf{p}} - \frac{1}{8m^3}\hat{\mathbf{p}}^4]. \qquad (B9)$$

Evaluating this in the usual way gives

$$\dot{\hat{z}} = \frac{1}{2m}\left[\hat{p}_z(1+3\Phi_G) + (1+3\Phi_G)\hat{p}_z\right] - \frac{1}{2}\left(\frac{\hat{\mathbf{p}}^2}{m^2}\right)\frac{\hat{p}_z}{m} \qquad (B10)$$

It is worth noting that $\Phi_G$ appears directly in the expression for the velocity.

The acceleration in the z direction is then given by

$$\ddot{\hat{z}} = \frac{1}{i}[\dot{\hat{z}}, \hat{H}]. \qquad (B11)$$

Using the same techniques, it is straightforward but somewhat tedious to show that

$$\ddot{\hat{z}} = -\nabla\Phi_G\left[1 + \frac{\hat{\mathbf{p}}_x^2 + \hat{\mathbf{p}}_y^2 - 3\hat{\mathbf{p}}_z^2}{m^2}\right]. \qquad (B12)$$

Taking the expectation value of both sides of Eq. (B12) gives

$$\ddot{z} = -\nabla\Phi_G\left[1 + v_x^2 + v_y^2 - 3v_z^2\right]. \qquad (B13)$$

The dependence of the velocity on $\Phi_G$ in Eq. (B10) would give a contribution that is second-order in $\Phi_G$ and is neglected here.

The acceleration of a classical particle in a uniform gravitational field is proportional to the Christoffel symbol that appears in the covariant derivative of a vector. It is interesting to note that the acceleration of an electron in a uniform gravitational field in the Dirac theory is due to the dependence of the $\gamma^\mu$ term on $\Phi_G(z)$, while the contribution from $\Gamma_\mu$ in the covariant derivative for a spinor is negligible here as is shown in a subsequent section.

The analytic calculations were first performed by hand and then repeated using the symbolic manipulation capabilities of Mathematica, which gave the same results. The acceleration was also calculated analytically using a different method based on a series of transformations, which also gave the same results but is too lengthy to describe here.

## APPENDIX C: DETAILS OF THE NUMERICAL CALCULATIONS

The full Dirac equation (12) with the $\Gamma_\mu$ term included was integrated numerically using Mathematica. This approach made no approximations other than $\Phi_G/c^2 \ll 1$. More general calculations were also performed in which $\Phi_G/c^2$ was not assumed to be small. The results of those calculations were also in agreement with the classical geodesic described in Appendix F although the results will not be described in detail here.

The initial values of $\phi_1$ and $\chi_1$ were chosen to correspond to a superposition of positive-energy plane wave solutions. This was accomplished by first choosing a Gaussian wave function for $\phi_1$ and then taking its Fourier transform to determine its momentum components. The corresponding positive-energy plane-wave momentum components [16] for $\chi_1$ were then calculated and their inverse Fourier transform was used as the initial value of $\chi_1$. This ensured that the initial state corresponded to an electron with no positron components and it eliminated any zitterbewegung motion.

The integration was performed using the NDSolve routine of Mathematica, which can automatically adjust the time increment and spatial grid size to achieve the desired accuracy. In practice, a fixed grid containing $1.2\times 10^7$ spacetime points was used for all the calculations in order to eliminate any slight dependence of the results on the grid spacing. It was found that the use of the fixed Runge-Kutta algorithm within NDSolve gave the best stability and precision.

The calculations were performed on a work station with dual processors, each of which had four cores. Mathematica can make use of parallel processing on all eight cores and the calculations typically required less than 15 minutes each. Up to 25 gigabytes of random access memory was required, however. The numerical results were estimated to be correct to approximately 4 significant digits based on the observed dependence of the results on changes in various parameters such as the grid size.

It was necessary to define a suitable measure of the position of a wave packet in order to calculate its acceleration. The mean location $\bar{z}$ of the wave packet was taken to be

$$\bar{z} = \frac{\int z \psi^\dagger \psi \, dz}{\int \psi^\dagger \psi \, dz}. \tag{C1}$$

The expectation value of z was normalized in this way because the Dirac equation (3) is not Hermitian in general. The value of $\psi^\dagger \psi$ corresponds to the charge density rather than a probability density, but both components of the wave function are accelerated in the same direction (unlike the situation in an electromagnetic field.) It is apparent from Fig. 3 that the wave packet remains very nearly a Gaussian and that its position is well-defined. The calculated value of $\bar{z}$ was used to estimate the acceleration $a_z$ using the relation $\Delta \bar{z} = a(\Delta t)^2 / 2$ where $\Delta \bar{z}$ is the change in $\bar{z}$ over a time interval of $\Delta t$.

The main goal of the numerical calculations was to confirm the results of the analytic calculations and to provide some additional insight such as in Figs. 2 and 3. The predicted acceleration of an electron is independent of its mass as can be seen from Eq. (17). Thus there was no reason to use the actual mass of the electron and $m = 100$ was arbitrarily chosen (in units with $\hbar = c = 1$) to minimize the rapid phase oscillations at a frequency of $mc^2 / \hbar$. The mass was chosen to be sufficiently large that there were a large number of phase oscillations over the time period of interest, however.

The numerical data points shown in Fig. 4 were all calculated using a relatively small value of $\nabla \Phi = 0.001$ in these units. This ensured that the change in velocity due to the gravitational acceleration was negligible compared to the initial velocity, so that each data point corresponded to a well-defined velocity. It can be seen that the numerical results were in agreement with the analytic calculations to order $v^2 / c^2$.

**APPENDIX D: SPIN CONNECTION $\Gamma_\mu$**

It will now be shown that the spin connection $\Gamma_\mu$ in the covariant derivative of Eq. (5) can be neglected for a uniform gravitational field, which is the situation of interest here.

It is possible to construct a local inertial reference frame $\xi_X^\alpha$ at every point $X$ in spacetime in such a way that the metric $g^{\mu\nu} = \eta^{\mu\nu}$ and its first derivatives vanish at the origin. For the uniform gravitational field illustrated in Fig. 1, a local inertial frame $\xi_X^\alpha(x)$ centered at the point $X$ can be described in terms of the global coordinates $x^\mu$ by the transformation

$$\xi_X^0 = (1+\Phi_0)\tilde{x}^0 + \nabla\Phi_G \tilde{x}^0 \tilde{x}^3$$
$$\xi_X^1 = (1-\Phi_0)\tilde{x}^1 - \nabla\Phi_G \tilde{x}^1 \tilde{x}^3$$
$$\xi_X^2 = (1-\Phi_0)\tilde{x}^2 - \nabla\Phi_G \tilde{x}^2 \tilde{x}^3$$
$$\xi_X^3 = (1-\Phi_0)\tilde{x}^3 + \frac{1}{2}\nabla\Phi_G \left(\tilde{x}^1\tilde{x}^1 + \tilde{x}^2\tilde{x}^2 - \tilde{x}^3\tilde{x}^3 + \tilde{x}^0\tilde{x}^0\right).$$
(D1)

Here $\tilde{x}^\mu \equiv x^\mu - X^\mu$ is the distance from point $X$, $\Phi_0 = \Phi_G(X)$, and $\nabla\Phi_G$ is a constant. It is straightforward to show that Eq. (D1) gives $g^{\mu\nu} = \eta^{\mu\nu}$ at $X = 0$ with vanishing first derivatives as required.

From Eqs. (A13) and (D1), the nonzero components of the vierbein are given by

$$V^0_{\ 0} = 1 + \Phi_0$$
$$V^1_{\ 1} = V^2_{\ 2} = V^3_{\ 3} = 1 - \Phi_0. \tag{D2}$$

The nonzero values of the Christoffel symbols [12] for the metric of Eq. (7) are

$$\Gamma^0_{\ 03} = \Gamma^0_{\ 30} = \nabla\Phi_G$$
$$\Gamma^1_{\ 13} = \Gamma^1_{\ 31} = -\nabla\Phi_G$$
$$\Gamma^2_{\ 23} = \Gamma^2_{\ 32} = -\nabla\Phi_G \tag{D3}$$
$$\Gamma^3_{\ 00} = \Gamma^3_{\ 11} = \Gamma^3_{\ 22} = \nabla\Phi_G$$
$$\Gamma^3_{\ 33} = -\nabla\Phi_G.$$

Combining Eqs. (A8), (A17), (D2), and (D3) gives

$$\Gamma_0 = \frac{1}{2}\nabla\Phi_G \alpha_{(0)}^3$$
$$\Gamma_1 = \frac{i}{2}\nabla\Phi_G \sigma_y$$
$$\Gamma_2 = -\frac{i}{2}\nabla\Phi_G \sigma_x \tag{D4}$$
$$\Gamma_3 = 0.$$

Here $\sigma_i$ is a $4 \times 4$ block-diagonal matrix with $\tau_i$ on the diagonals.

Inserting $\Gamma_\mu$ into the Dirac equation produces an additional term $U$ in the Hamiltonian given by

$$U = \frac{i}{2} \nabla \Phi_G \alpha_{(0)}{}^3. \tag{D5}$$

The ratio $\varepsilon$ of the magnitude of this term to the usual gravitational potential $m\Phi_G$ is on the order of

$$\varepsilon = \frac{U}{m\Phi_G} \sim \frac{v}{c} \frac{\Phi_G}{R_0} \frac{1}{m\Phi_G} = \frac{v}{c} \left(\frac{h}{mc}\right) \frac{1}{R_0} = \frac{v}{c} \frac{\lambda_C}{R_0}. \tag{D6}$$

Here $\lambda_C = h/mc$ is the Compton radius of an electron $(2.4 \times 10^{-10} \text{ cm})$ and we have reinserted the constants $\hbar$ and $c$ in order to make this connection. $R_0$ is the distance to the source of the gravitational field which is many orders of magnitude larger than $\lambda_C$. Eq. (D6) shows that the $\Gamma_\mu$ term is negligible here, although it was included in the numerical calculations to avoid making any assumptions other than $\Phi_G/c^2 \ll 1$.

Aside from being extremely small, the spin connection $\Gamma_\mu$ does not produce any acceleration in a uniform gravitational field because it is independent of the position and has no gradient. It does produce a small spin-dependent acceleration and a departure from a geodesic in a non-uniform gravitational field [4-9]. The same thing occurs for a rotating classical object due to tidal forces and there is no violation of the equivalence principle in that case [8].

**APPENDIX E: CONNECTION WITH THE FOLDY-WOUTHUYSEN TRANSFORMATION**

Wave packets can have a significant spatial extent, in which case the position of a particle and thus its acceleration are intrinsically uncertain to some extent. Nevertheless, we can consider a situation in which $R_0$ is so large that we can take

$$\lambda_d \ll d \ll \Delta \bar{z} \ll R_0. \tag{E1}$$

Here $\lambda_d$ is the De Broglie wavelength, $d$ is a typical dimension of the wave packet, and $\Delta \bar{z}$ is the displacement of the wave packet due to the acceleration of gravity. Requiring $\lambda_d \ll d$ ensures that dispersion will not be significant, $d \ll \Delta \bar{z}$ ensures that the position of the wave packet is well-defined on the scale of the motion, and $\Delta \bar{z} \ll R_0$ ensures that the gravitational field is uniform over these distances. The position and acceleration of an electron can be arbitrarily well defined in this limit. The analytic calculations described in the text can be applied in this limit and the expectation values taken in going from Eq. (16) to (17) provide a valid measure of the acceleration of an electron. Similar comments apply to the definition of the location $\bar{z}$ of the wave packet in Eq. (C1) that was used in the numerical calculations.

In special relativity, the operator $\hat{x}$ is sometimes replaced with another operator $\hat{X}$ using the Foldy-Wouthuysen transformation [14]. This is designed to eliminate the zitterbewegung motion that occurs when there is interference between positive and negative energy components (electrons and positrons in a second-quantized description). This does not appear to be a problem here because the initial state was chosen to contain only positive-energy components and the gravitational field accelerates both components in the same way [17].

Nevertheless, it can be shown that the Foldy-Wouthuysen transformation gives the same results as the iterative approach discussed above. Here an appropriate unitary transformation $\hat{U}$ is applied to both the wave function and the Hamiltonian:

$$\begin{aligned} \psi' &= \hat{U}\psi \\ \hat{H}' &= \hat{U}\hat{H}\hat{U}^\dagger. \end{aligned} \tag{E2}$$

In the non-relativistic limit, Obukhov [9] showed that this gives

$$\hat{H}' = \frac{1}{4m}\left(W^{-1}\hat{\mathbf{p}}^2 F + F\hat{\mathbf{p}}^2 W^{-1}\right) + Vmc^2, \tag{E3}$$

where we have not included the spin-dependent terms that vanish in a uniform gravitational field. For the Schwarzschild metric, the functions $F$, $V$, and $W$ are defined [9] by

$$\begin{aligned} V &= \left(1 - \frac{GM}{2c^2 R}\right)\left(1 + \frac{GM}{2c^2 R}\right)^{-1} \\ W &= \left(1 + \frac{GM}{2c^2 R}\right)^2 \\ F &= V/W. \end{aligned} \tag{E4}$$

The Hamiltonian of Eq. (E3) corresponds to the first term in a Taylor series expansion. It is straightforward to include the second-order term in Eq. (E3) as well, which gives

$$H' = \frac{1}{4m}\left[(1+\Phi_G)\hat{\mathbf{p}}^2(1+2\Phi_G) + (1+2\Phi_G)\hat{\mathbf{p}}^2(1+\Phi_G)\right] \\ + (1+\Phi_G)mc^2 - \frac{1}{8}\frac{\hat{\mathbf{p}}^4}{m^3 c^2}. \tag{E5}$$

Here the functions $F$, $V$, and $W$ in Eq. (E4) have been evaluated to first order in $\Phi_G$ and higher order terms such as $\hat{p}^4 \Phi_G \psi$ have been dropped.

By comparing the terms involving $\nabla \Phi_G$, it can be seen that the Hamiltonian $\hat{H}'$ from the Foldy-Wouthuysen transformation in Eq. (E5) is identical to the Hamiltonian $\hat{H}$ obtained from the iterative approach in Eq. (13), aside from the spin-dependent terms. The time dependence of $\hat{X}$ obtained using commutator techniques would therefore give the same acceleration as that obtained earlier in Eq. (17). It should be noted that Obukhov [9] considered only the spin-dependent effects in the nonrelativistic limit and not the velocity-dependent acceleration.

Kono and Kasai [2] considered the spin-dependent departure of an electron from a geodesic due to the $\Gamma_\mu$ term in the Dirac equation in the extreme relativistic limit. However, they simply assumed that the "baseline" trajectory in the absence of spin-dependent effects was a geodesic and calculated small spin-dependent departures from it. A similar assumption was made in Ref. [8].

## APPENDIX F: CLASSICAL GEODESIC

The classical acceleration of a particle along a geodesic trajectory will be calculated in this appendix for comparison with the results from the Dirac equation in curved spacetime. Here we will use a technique based on a Taylor series expansion over an infinitesimal proper time interval $\Delta \tau$, although there are other ways of doing this as well.

The geodesic equation is given [15] by

$$\frac{d^2 x^\mu}{d\tau^2} = -\Gamma^\mu_{\alpha\beta} \frac{dx^\alpha}{d\tau} \frac{dx^\beta}{d\tau}. \tag{F1}$$

The nonzero elements of $\Gamma^\mu_{\alpha\beta}$ are given to lowest-order in $\Phi_G$ in Eq. (D3) for the case in which the source of the gravitational field is far from the point of interest. Here we will use the exact values of $\Gamma^\mu_{\alpha\beta}$ for the Schwarzschild metric and then take the limit of a distant source at the end to obtain the acceleration in a uniform gravitational field.

Eq. (F1) gives the second derivatives with respect to the proper time $\tau$ but we need to calculate the derivatives with respect to the time itself in order to determine the acceleration. In order to do this, we can consider an infinitesimal proper time interval $\Delta \tau$ and expand the coordinates of the particle in a Taylor series about an initial value of $\tau = \tau_0$:

$$z(\tau) = z_0 + \left[\frac{dz}{d\tau}\right]_0 \Delta\tau + \frac{1}{2}\left[\frac{d^2 z}{d\tau^2}\right]_0 \Delta\tau^2$$
$$t(\tau) = t_0 + \left[\frac{dt}{d\tau}\right]_0 \Delta\tau + \frac{1}{2}\left[\frac{d^2 t}{d\tau^2}\right]_0 \Delta\tau^2. \tag{F2}$$

Here the subscript 0 indicates that these derivatives are to be evaluated at the initial proper time $\tau_0$, where $z$ and $t$ have the values $z_0$ and $t_0$, respectively. Without loss of generality, we can take $z_0 = t_0 = \tau_0 = 0$.

It will be convenient to define $f_0$ as

$$f_0 = \left[\frac{dt}{d\tau}\right]_0 = 1 / \left[\sqrt{g_{\mu\nu} \frac{dx^\mu}{dt} \frac{dx^\nu}{dt}}\right]_0. \tag{F3}$$

The right-hand side of Eq. (F3) follows directly from the definition of the proper time in terms of the metric. For the Schwarzschild metric, $f_0$ has the value

$$f_0 = \frac{1}{\sqrt{(1+2\Phi_0) - (1-2\Phi_0) v_{z0}^2}}, \tag{F4}$$

where $\Phi_0$ is $\Phi_G$ evaluated at $z_0$ and $t_0$. From the chain rule we can then write

$$\left[\frac{dz}{d\tau}\right]_0 = \left[\frac{dz}{dt}\right]_0 \left[\frac{dt}{d\tau}\right]_0 = f_0 v_{z0}. \tag{F5}$$

Eq. (F1) can be used to obtain

$$\left[\frac{d^2 z}{d\tau^2}\right]_0 = -\frac{\nabla \Phi_G}{(1-2\Phi_0)}\left[\frac{dt}{d\tau}\right]_0^2 + \frac{\nabla \Phi_G}{(1-2\Phi_0)}\left[\frac{dz}{d\tau}\right]_0^2, \tag{F6}$$

where it is understood that $\nabla \Phi_G$ is to be evaluated at $\tau_0 = 0$. Making use of the definition of $f_0$ allows this to be rewritten as

$$\left[\frac{d^2 z}{d\tau^2}\right]_0 = -\frac{\nabla \Phi_G}{(1-2\Phi_0)} f_0^2 (1 - v_{z0}^2). \tag{F7}$$

Proceeding in the same way for the time derivatives gives

$$\left[\frac{d^2 t}{d\tau^2}\right]_0 = -2 \frac{\nabla \Phi_G}{(1+2\Phi_0)} f_0^2 v_{z0}. \tag{F8}$$

Inserting Eqs. (F7) and (F8) back into the Taylor series expansion of Eq. (F2) gives

$$z(\tau) = f_0 v_{z0}\tau - \frac{1}{2}\frac{\nabla\Phi_G}{(1-2\Phi_0)}f_0^2(1-v_{z0}^2)\tau^2 \quad (F9)$$

and

$$t(\tau) = f_0\tau - \frac{\nabla\Phi_G}{(1+2\Phi_0)}f_0^2 v_{z0}\tau^2. \quad (F10)$$

Here we have used $\Delta\tau = \tau$ for $\tau_0 = 0$.

Eq. (F10) can now be solved to express $\tau$ as a function of $t$. From the quadratic formula, the result is

$$\tau(t) = \frac{f_0 - f_0\sqrt{1-4\nabla\Phi_G v_{z0}/(1+2\Phi_0)t}}{\left[2f_0^2\nabla\Phi_G v_{z0}/(1+2\Phi_0)\right]}. \quad (F11)$$

With $t_0 = 0$ and an infinitesimal $\Delta\tau$, the square root can be expanded in a Taylor series to second order in $t$ to obtain

$$\tau(t) = \frac{1}{f_0}\left[t + \frac{\nabla\Phi_G}{(1+2\Phi_0)}v_{z0}t^2\right]. \quad (F12)$$

Inserting this expression for $\tau(t)$ back into Eq. (F9) gives z as a function of $t$, which can be differentiated to give

$$\frac{d^2z}{dt^2} = -\frac{\nabla\Phi_G}{(1-4\Phi_G^2)}\left[(1+2\Phi_G)-3v_{z0}^2\left(1-\frac{2}{3}\Phi_G\right)\right]. \quad (F13)$$

Here we have replaced $\Phi_0$ with $\Phi_G$ as $\Delta\tau \to 0$. Eq. (F13) gives the acceleration of a particle as observed in a reference frame with the Schwarzschild metric. This result reduces to the acceleration shown in Eq. (18) in the text in the limit of $\Phi_G/c^2 \ll 1$.